\documentclass[preprint,10pt]{aastex}

\usepackage{graphicx}
\usepackage{epsfig}
\usepackage{multirow}
\usepackage{stmaryrd}

\def\hi{\relax \ifmmode {\rm H\,{\sc i}}\else H\,{\sc i}\fi}
\def\arcsec{\hbox{$^{\prime\prime}$}}

\shorttitle{The unusual vertical distribution of of NGC~4013}
\shortauthors{Comer\'on et al.}

\begin{document}

\title{The unusual vertical mass distribution of NGC~4013 seen through the Spitzer Survey of Stellar Structure in Galaxies (S$^4$G)}

\author{
S\'ebastien Comer\'on,\altaffilmark{1}
Bruce G.~Elmegreen,\altaffilmark{2}
Johan H.~Knapen,\altaffilmark{3,4}
Kartik Sheth,\altaffilmark{5}
Joannah L.~Hinz,\altaffilmark{6}
Michael W.~Regan,\altaffilmark{7}
Armando Gil de Paz,\altaffilmark{8}
Juan-Carlos Mu\~noz-Mateos,\altaffilmark{5}
Kar\'in Men\'endez-Delmestre,\altaffilmark{9}
Mark Seibert,\altaffilmark{9}
Taehyun Kim,\altaffilmark{5}
Trisha Mizusawa,\altaffilmark{5}
Eija Laurikainen,\altaffilmark{10,11}
Heikki Salo,\altaffilmark{10}
Jarkko Laine,\altaffilmark{10}
E.~Athanassoula,\altaffilmark{12}
Albert Bosma,\altaffilmark{12}
Ronald J.~Buta,\altaffilmark{13}
Dimitri A.~Gadotti,\altaffilmark{14}
Luis C.~Ho,\altaffilmark{9}
Benne Holwerda,\altaffilmark{15,16}
Eva Schinnerer\altaffilmark{17}
and Dennis Zaritsky\altaffilmark{6}}

\altaffiltext{1}{Korea Astronomy and Space Science Institute, 776 Daedeokdae-ro, Yuseong-gu, Daejeon 305-348, Republic of Korea}
\altaffiltext{2}{IBM T.~J.~Watson Research Center, 1101 Kitchawan Road, Yorktown Heights, NY 10598, USA}
\altaffiltext{3}{Instituto de Astrof\'isica de Canarias, E-38200 La Laguna, Spain}
\altaffiltext{4}{Departamento de Astrof\'isica, Universidad de La Laguna, E-38205 La Laguna, Tenerife, Spain}
\altaffiltext{5}{National Radio Astronomy Observatory / NAASC, 520, Edgemont Road, Charlottesville, VA 22903, USA}
\altaffiltext{6}{Steward Observatory, University of Arizona, 933 North Cherry Avenue, Tucson, AZ 85721, USA}
\altaffiltext{7}{Space Telescope Science Institute, 3700 San Martin Drive, Baltimore, MD 21218, USA}
\altaffiltext{8}{Departamento de Astrof\'isica, Universidad Complutense de Madrid, 28040, Madrid Spain}
\altaffiltext{9}{The Observatories of the Carnegie Institution of Washington, 813 Santa Barbara Street, Pasadena, CA 91101, USA}
\altaffiltext{10}{Astronomy Division, Department of Physical Sciences, P.~O.~Box 3000, FIN-90014 University of Oulu, Finland}
\altaffiltext{11}{Finnish Centre of Astronomy with ESO (FINCA), University of Turku, V\"ais\"al\"antie 20, FI-21500, Piikki\"o, Finland}
\altaffiltext{12}{Laboratoire d'Astrophysique de Marseille (LAM), UMR6110, Universit\'e de Provence/CNRS, Technop\^ole de Marseille \'Etoile, 38 rue Fr\'ed\'eric Joliot Curie, 13388 Marseille C\'edex 20, France}
\altaffiltext{13}{Department of Physics and Astronomy, University of Alabama, Box~870324, Tuscaloosa, AL~35487, USA}
\altaffiltext{14}{European Southern Observatory, Casilla 19001, Santiago 19, Chile}
\altaffiltext{15}{European Space Agency, ESTEC, Keplerlaan 1, 2200, AG, Noordwijk, the Netherlands}
\altaffiltext{16}{Astrophysics, Cosmology and Gravity Centre (AC$\lightning$GC)}
\altaffiltext{17}{Max-Planck-Institut f\"ur Astronomie, K\"onigstuhl 17, 69117 Heidelberg, Germany}

\begin{abstract}

NGC~4013 is a nearby Sb edge-on galaxy known for its ``prodigious'' \hi\ warp and its ``giant'' tidal stream. Previous work on this unusual object shows that it cannot be fitted satisfactorily by a canonical thin+thick disk structure. We have produced a new decomposition of NGC~4013, considering three stellar flattened components (thin+thick disk plus an extra and more extended component) and one gaseous disk. All four components are considered to be gravitationally coupled and isothermal. To do so, we have used the $3.6\mu{\rm m}$ images from the Spitzer Survey of Stellar Structure in Galaxies (S$^4$G).

We find evidence for NGC~4013 indeed having a thin and a thick disk and an extra flattened component. This smooth and extended component (scaleheight $z_{\rm EC}\sim3$\,kpc) could be interpreted as a thick disk or as a squashed ellipsoidal halo and contains $\sim20\%$ of the total mass of all three stellar components. We argue it is unlikely to be related to the ongoing merger or due to the off-plane stars from a warp in the other two disk components. Instead, we favor a scenario in which the thick disk and the extended component were formed in a two-stage process, in which an initially thick disk has been dynamically heated by a merger soon enough in the galaxy history to have a new thick disk formed within it.

\end{abstract}


\section{Introduction}

Thick disks, first detected by Burstein (1979) and Tsikoudi (1979), are seen in edge-on galaxies as excesses of light a few thin disk scaleheights above the galaxy mid-planes. They are known to be ubiquitous (Yoachim \& Dalcanton 2006; Comer\'on et al.~2011a) and their properties give important clues for understanding galaxy formation and evolution (Comer\'on et al.~2011b and references therein; hereafter CO11b). Recent studies (Robertson et al.~2006; Elmegreen \& Elmegreen 2006; Brook et al.~2007; Richards et al.~2010) suggest an in situ formation mechanism for a significant fraction of the thick disk mass. This was recently supported further by our result that the masses of thick and thin disks are of the same order (CO11b). Other formation mechanisms such as disk kinematical heating due to its own overdensities (Villumsen 1985; H\"anninen \& Flynn 2002; Sch\"onrich \& Binney 2009; Bournaud et al.~2009) and the accretion of stars from infalling satellites (Statler 1988; Gilmore et al.~2002; Abadi et al.~2003; Navarro et al.~2004; Martin et al.~2004; Read et al.~2008) also contribute to the thick disk mass (CO11b).

In CO11b we made thin+thick disk decompositions of 46 edge-on galaxies using images from the Spitzer Survey of Stellar Structure in Galaxies (S$^4$G; Sheth et al 2010). However, two galaxies, ESO~079-003 and NGC~4013, could not be successfully fitted down to low surface brightness due to the presence of an extra light component not accounted for in our fits [affecting significantly the luminosity profiles starting at $\mu_{3.6\mu{\rm m}}({\rm AB})=23\,{\rm mag\,arcsec^{-2}}$]. In CO11b, NGC~3628 also presents an obvious third component, but at a lower surface brightness [$\mu_{3.6\mu{\rm m}}({\rm AB})=24.5\,{\rm mag\,arcsec^{-2}}$]. The aim of this Letter is to study the properties of these extra components. As ESO~079-003 is located near a bright star which makes its study at low surface brightness levels difficult and NGC~3628 presents a disturbed morphology due to a recent interaction we focused on NGC~4013.

NGC~4013 is an Sb galaxy (Buta et al.~2007), at $18.6\pm2.5$\,Mpc (NED average for 13 redshift-independent measurements). Its optical diameter is $D_{25}=294\arcsec$ (HyperLEDA; Paturel et al.~2003). It is often described using superlative adjectives: it has a ``prodigious'' \hi\ warp (Bottema et al.~1987; Bottema 1995; 1996) and a ``giant'' stellar tidal stream with an age of a few Gyr (Mart\'inez-Delgado et al.~2009; hereafter MD09). The \hi\ warp starts just at the optical edge of the galaxy (Bottema~1995) and it is one of the largest warps ever observed. It may have been triggered by the minor merger event which caused the tidal stream (MD09). In addition, NGC~4013 has a boxy bulge (Jarvis 1986) which could be tracing an edge-on bar (Athanassoula \& Misiriotis 2002; Mart\'inez-Valpuesta et al.~2006) or be the consequence of a merging event (Binney \& Petrou 1985).                                                    

\begin{figure}
\begin{center}
\begin{tabular}{c}
\includegraphics[width=0.45\textwidth]{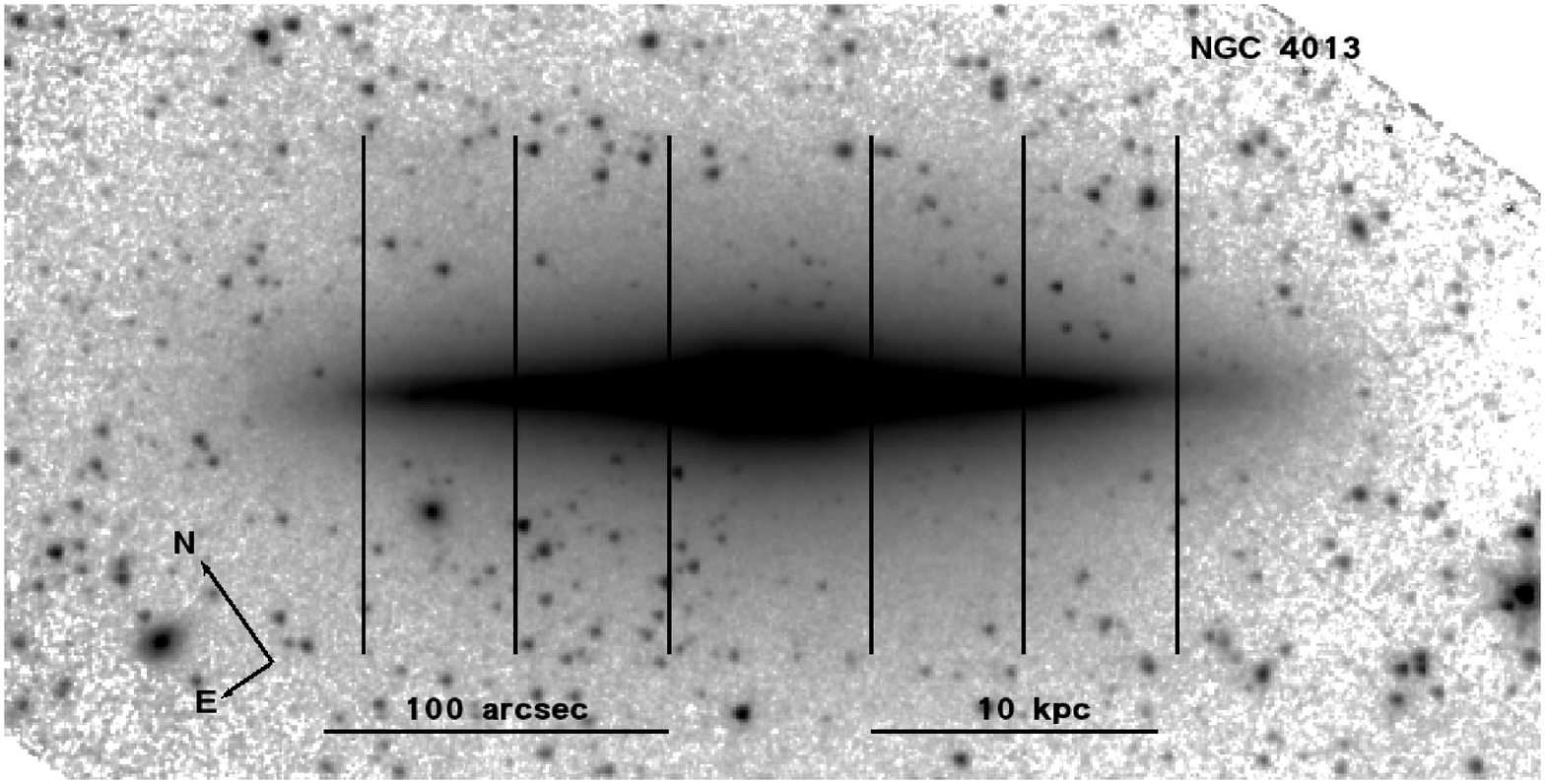}\\
\includegraphics[width=0.45\textwidth]{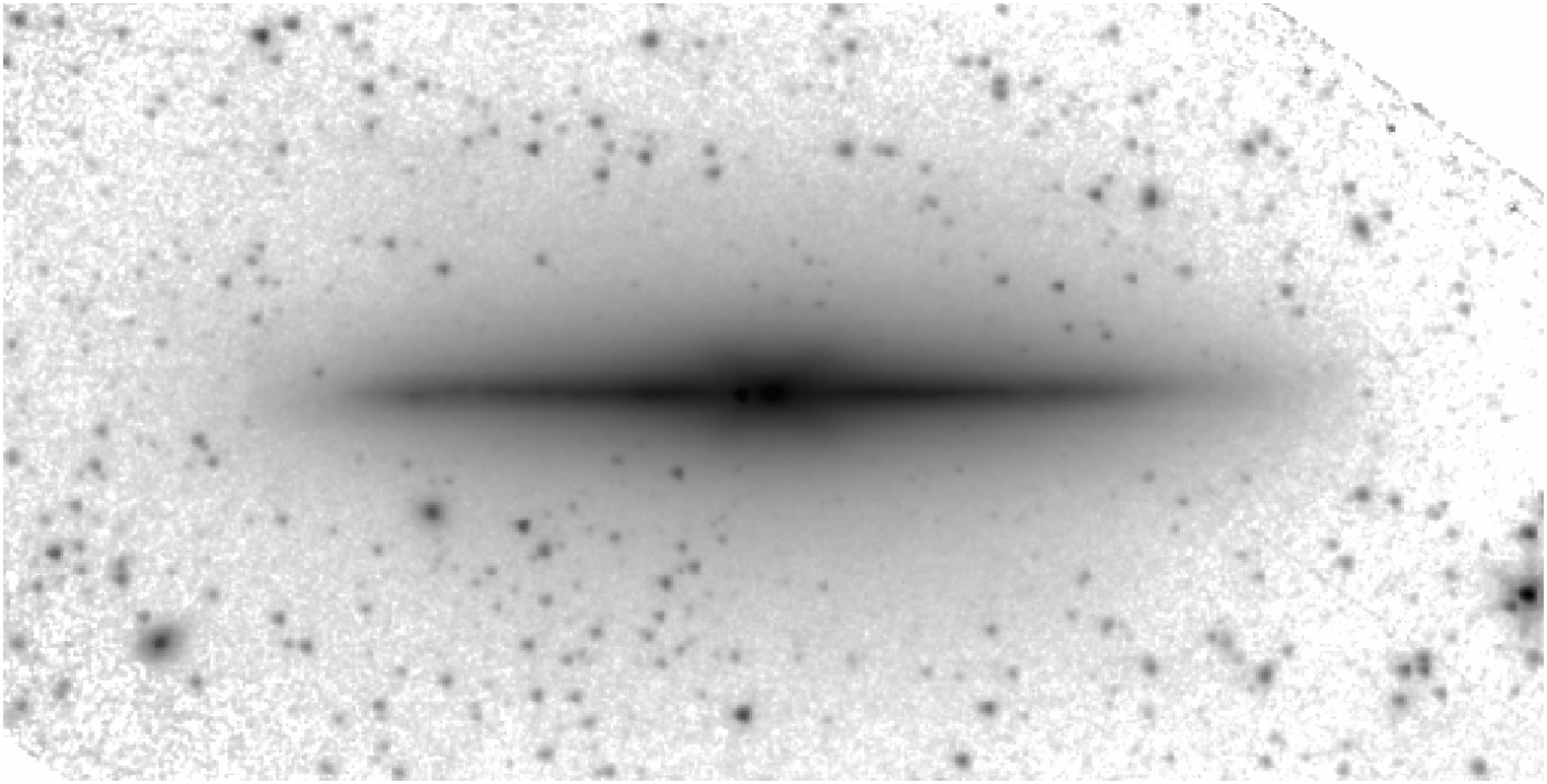}\\
\end{tabular}
\caption{\label{NGC4013} $3.6\mu{\rm m}$-band S$^4$G image of NGC~4013 with two different luminosity stretches. The vertical lines indicate the limits of the bins for which luminosity profiles have been produced, from left to right, at galactocentric distances of $-0.8\,r_{25}<R<-0.5\,r_{25}$, $-0.5\,r_{25}<R<-0.2\,r_{25}$, $-0.2\,r_{25}<R<0.2\,r_{25}$, $0.2\,r_{25}<R<0.5\,r_{25}$ and $0.5\,r_{25}<R<0.8\,r_{25}$. In order to avoid the influence of the bulge we have not produced fits for the central bin.}
\end{center}
\end{figure}

The S$^4$G $3.6\,\mu{\rm m}$-band image used for the study presented in this Letter is shown in Fig.~\ref{NGC4013}.

\section{Fitting procedure}

In CO11b, the observed luminosity profiles were fitted with synthetic profiles resulting from coupling two stellar plus one gaseous isothermal disks (using the equations in Narayan \& Jog 2002). The solutions are not analytic, so the fit was done by comparing the luminosity profile to a grid of pre-computed models with different thick-to-thin central mass density ratios ($\rho_{\rm T0}/\rho_{\rm t0}$) and thick-to-thin velocity dispersions in the vertical direction ratios ($\sigma_{\rm T}/\sigma_{\rm t}$) where $T$ stands for the thick disk and $t$ stands for the thin disk. Both observed luminosity and pre-computed profiles were scaled to have a mid-plane luminosity equal to unity and to have $\rho(z=200)/\rho(z=0)=0.1$. In addition the synthetic profiles were convolved with a Gaussian kernel with a full width at half maximum (FWHM) equal to 2.2\arcsec\ in order to account for the point spread function of the $3.6\mu{\rm m}$-band image (S$^4$G ``super-PSF''; Sheth et al.~2010). The mass-to-light ratios ($\Upsilon$) for both thin and thick disk stars had to be assumed in order to make the computation of the grid of models, and reasonable limiting cases with $\Upsilon_{\rm T}/\Upsilon_{\rm t}=1.2$ and $\Upsilon_{\rm T}/\Upsilon_{\rm t}=2.4$ were studied (see CO11b). The gaseous disk was assumed to have a column mass density equal to a 20\% of that of the thin disk. The case without gas was also studied yielding similar results to the with-gas case, but the fit quality was slightly worse.

The fits were made ignoring the mid-plane pixel and over a range of magnitudes, $\Delta m$, which was defined as the range for which either the square root of the $\chi^2$ of the fit was smaller than $\sqrt{\chi^2}<0.1\,{\rm mag\,arcsec^{-2}}$ or down to a limiting magnitude $\mu_{\rm l}({\rm AB})=26\,{\rm mag\,arcsec^{-2}}$. The discussion on how we chose this goodness-of-fit parameter and this limiting magnitude appear in CO11b.

In CO11b, for the NGC~4013 luminosity profiles at $0.2\,r_{25}<|R|<0.5\,r_{25}$ the fits could only be performed down to $\mu({\rm AB})\sim23\,{\rm mag\,arcsec^{-2}}$, at which point our criterion for the quality of the fit, $\sqrt{\chi^2}>0.1\,{\rm mag\,arcsec^{-2}}$, was not met. The reason for that was the presence of a third component which made our two stellar disk decomposition insufficient at describing NGC~4013. This component was called ``halo'' by van der Kruit \& Searle (1982) and by MD09 who reported that it is box-shaped (although the boxiness is probably caused by the location of the tidal streams and not by the intrinsic shape of the component).

We adapted the code used in C011b to include a third component in equilibrium with the two stellar disks and the gaseous disk. This is a natural approach, because the luminosity profile decays exponentially with height once in the range of heights for which the more extended component dominates, and because the solution of the equations for a set of coupled flattened components can be approximated as exponential at high distances above the mid-plane. However, the fact of adding a third flattened component does not imply that it is a disk; this is because, without kinematical information, a squashed ellipsoidal halo nature for this component cannot be discarded. Arguments in favor of this component being a disk are (i) its relatively high surface brightness, and (ii) its luminosity profile parallel to the mid-plane direction (not shown here) which is that of a typical edge-on disk, with a shallow slope for low $R$ and an exponential slope at larger $R$. Arguments in favor of the extended component being an squashed elliptical halo are (i) its isophotes are not disky (MD09) and may be very close to elliptical if the light of the tidal streams could be removed, and (ii) its ellipticity ($\epsilon=0.63$ when measured in an ellipse fit between radius 120\arcsec, which is the truncation radius of the thin disk, and 170\arcsec, where the extended component starts to be highly affected by noise) is compatible with that of simulated elliptical haloes (see, e.g., Lee et al.~2005). Due to its uncertain nature we will hereafter term this component as ``extended component'' (EC).

\begin{figure}
\begin{center}
\begin{tabular}{c}
\includegraphics[width=0.45\textwidth]{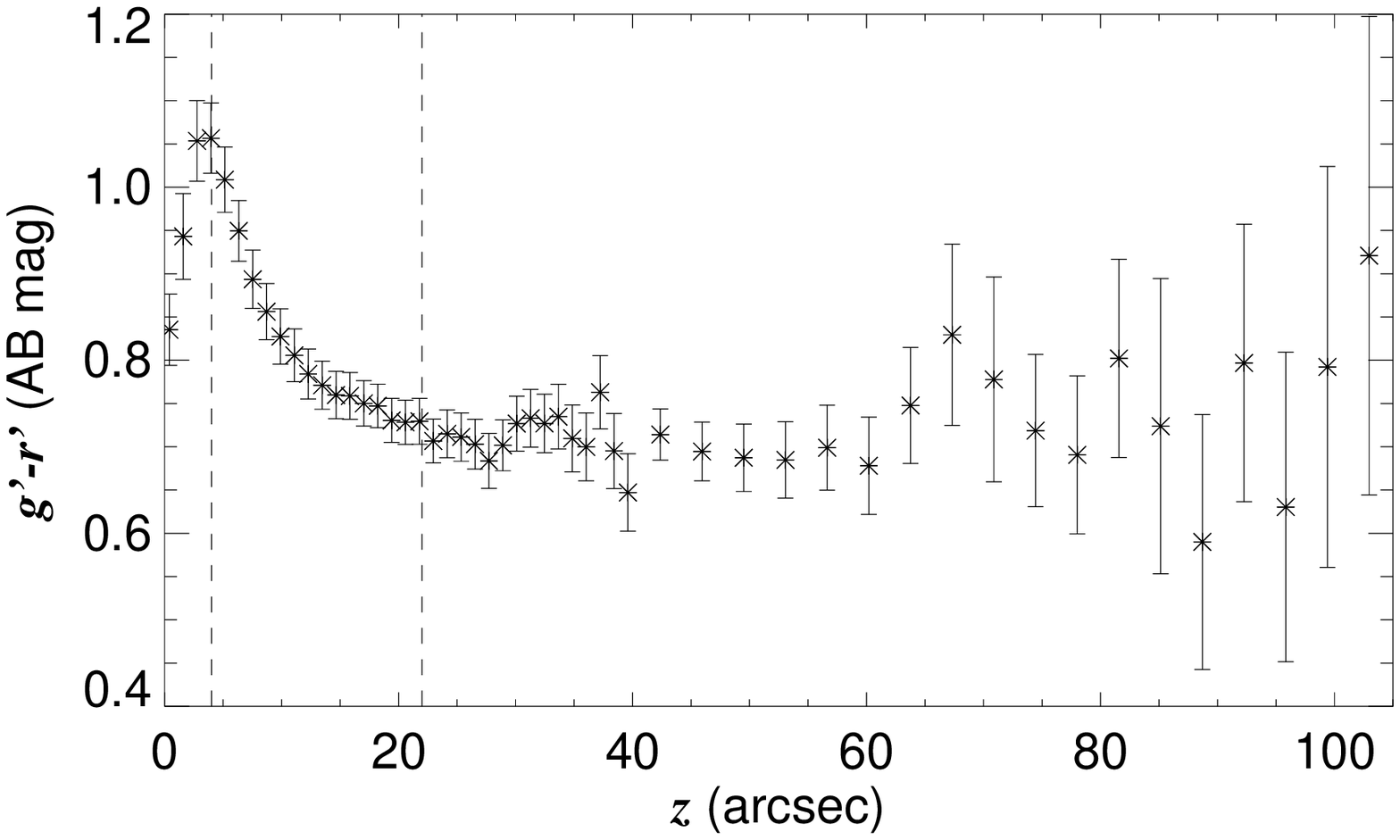}\\
\end{tabular}
\caption{\label{colors} $g'-r'$ color profiles made for $0.2\,r_{25}<|R|<0.5\,r_{25}$. The profile has been produced using SDSS DR7 data (Abazajian et al.~2007). Points for $z\leq40\arcsec$ are calculated every three pixels ($\sim1.2\arcsec$) and those with $z>40\arcsec$ are calculated every nine pixels ($\sim3.6\arcsec$). The error bars represent the $2\sigma$ statistical errors. The vertical lines separate the regions dominated by the luminosity of the thin disk, the thick disk and the EC for $\Upsilon_{\rm T}/\Upsilon_{\rm t}=1.2$.}
\end{center}
\end{figure}

The grid of models we used for the fits had $\rho_{\rm T0}/\rho_{\rm t0}$, $\rho_{\rm T0}/\rho_{\rm EC0}$, $\sigma_{\rm T}/\sigma_{\rm t}$ and $\sigma_{\rm T}/\sigma_{\rm EC}$ as free parameters, where $EC$ denotes the extended component. Our grid of models has been computed including a gaseous disk and using the same normalizations as in CO11b. We considered the cases $\Upsilon_{\rm T}/\Upsilon_{\rm t}=1.2$ and $\Upsilon_{\rm T}/\Upsilon_{\rm t}=2.4$. In both cases we set $\Upsilon_{\rm T}/\Upsilon_{\rm EC}=1.0$. This is justified by the fact that once above the mid-plane dust lane ($z>10\arcsec$, a region mainly influenced by the thick disk and the EC), the colors remain roughly constant, as can be seen in Fig.~\ref{colors}. Fig.~\ref{colors} provides extra evidence that the thick disk and the EC contain old star populations since the average color for $z>10\arcsec$ is $\overline{g'-r'(z>10\arcsec)}=0.73\,{\rm mag}$, which is compatible with the average colors predicted for S0 galaxies ($g'-r'=0.68\,{\rm mag}$; Fukugita et al.~1995).The reddening for $z<10\arcsec$ is caused by the mid-plane dust lane. Since we found no significant influence of this mid-plane dust on the $3.6\mu{\rm m}$-band profile for this galaxy in CO11b, we considered the mid-plane dust lane to have no effect on our fits.

\begin{figure*}
\begin{center}
\begin{tabular}{c}
\includegraphics[width=0.9\textwidth]{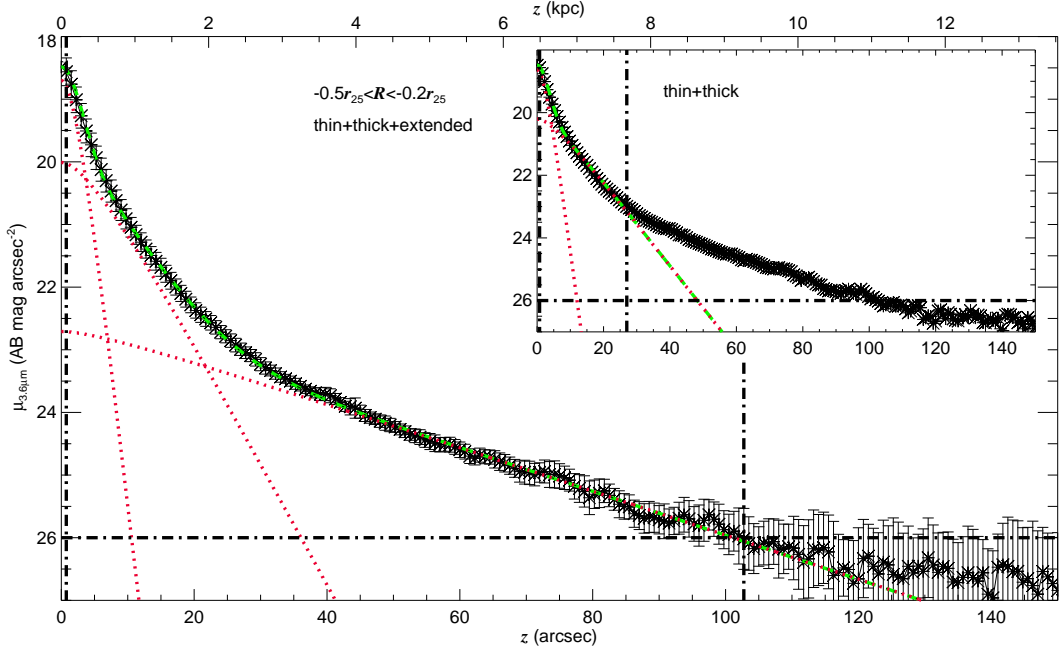}
\end{tabular}
\caption{\label{fits} Example of a fit to the vertical luminosity profiles of NGC~4013 considering $\Upsilon_{\rm T}/\Upsilon_{\rm t}=1.2$. Data points with error bars ($2\sigma$ statistical errors) represent the observed luminosity profile, and the dashed green curve the best fit. The dotted red curves indicate the contributions of the three stellar components. The dash-dotted vertical lines indicate the limits of the range in vertical distance above the mid-plane used for the fit. The horizontal line represents the $\mu({\rm AB})=26.0\,{\rm mag\,arcsec^{-2}}$ level down to which the fits have been done. The inset shows the thin+thick disk best fit obtained for NGC~4013 in CO11b.}
\end{center}
\end{figure*}

\begin{table}[t]
\begin{center}
\caption{\label{tablefit} Result of the fits.}
\begin{tabular}{l| c | c}
\hline
\hline
Fitting range & $\Upsilon_{\rm T}/\Upsilon_{\rm t}=1.2$ & $\Upsilon_{\rm T}/\Upsilon_{\rm t}=2.4$\\
\hline
\multirow{11}{*}{$-0.8\,r_{25}<R<-0.5\,r_{25}$}&\multicolumn{2}{c}{$\mu_{\rm l}=25.96\,{\rm mag\,arcsec^{-2}}$}\\
&\multicolumn{2}{c}{$\Delta m=6.0\,{\rm mag\,arcsec^{-2}}$}\\
&$\rho_{\rm T0}/\rho_{\rm t0}=0.35$&$\rho_{\rm T0}/\rho_{\rm t0}=0.64$\\
&$\rho_{\rm T0}/\rho_{\rm EC0}=7.00$&$\rho_{\rm T0}/\rho_{\rm EC0}=6.40$\\
&$\sigma_{\rm T}/\sigma_{\rm t}=2.55$&$\sigma_{\rm T}/\sigma_{\rm t}=2.69$\\
&$\sigma_{\rm T}/\sigma_{\rm EC}=0.38$&$\sigma_{\rm T}/\sigma_{\rm EC}=0.38$\\
&$\Sigma_{\rm T}/\Sigma_{\rm t}=1.21$&$\Sigma_{\rm T}/\Sigma_{\rm t}=2.23$\\
&$\Sigma_{\rm T}/\Sigma_{\rm EC}=1.54$&$\Sigma_{\rm T}/\Sigma_{\rm EC}=1.46$\\
&$z_{\rm t}=130$\,pc&$z_{\rm t}=120$\,pc\\
&$z_{\rm T}=610$\,pc&$z_{\rm T}=580$\,pc\\
&$z_{\rm EC}=3660$\,pc&$z_{\rm EC}=3500$\,pc\\
\hline
\multirow{11}{*}{$-0.5\,r_{25}<R<-0.2\,r_{25}$}&\multicolumn{2}{c}{$\mu_{\rm l}=25.99\,{\rm mag\,arcsec^{-2}}$}\\
&\multicolumn{2}{c}{$\Delta m=7.5\,{\rm mag\,arcsec^{-2}}$}\\
&$\rho_{\rm T0}/\rho_{\rm t0}=0.36$&$\rho_{\rm T0}/\rho_{\rm t0}=0.60$\\
&$\rho_{\rm T0}/\rho_{\rm EC0}=12.00$&$\rho_{\rm T0}/\rho_{\rm EC0}=12.00$\\
&$\sigma_{\rm T}/\sigma_{\rm t}=2.45$&$\sigma_{\rm T}/\sigma_{\rm t}=2.69$\\
&$\sigma_{\rm T}/\sigma_{\rm EC}=0.41$&$\sigma_{\rm T}/\sigma_{\rm EC}=0.41$\\
&$\Sigma_{\rm T}/\Sigma_{\rm t}=1.19$&$\Sigma_{\rm T}/\Sigma_{\rm t}=2.13$\\
&$\Sigma_{\rm T}/\Sigma_{\rm EC}=2.91$&$\Sigma_{\rm T}/\Sigma_{\rm EC}=3.02$\\
&$z_{\rm t}=110$\,pc&$z_{\rm t}=110$\,pc\\
&$z_{\rm T}=510$\,pc&$z_{\rm T}=530$\,pc\\
&$z_{\rm EC}=2780$\,pc&$z_{\rm EC}=2890$\,pc\\
\hline
\multirow{11}{*}{$0.2\,r_{25}<R<0.5\,r_{25}$}&\multicolumn{2}{c}{$\mu_{\rm l}=25.93\,{\rm mag\,arcsec^{-2}}$}\\
&\multicolumn{2}{c}{$\Delta m=7.5\,{\rm mag\,arcsec^{-2}}$}\\
&$\rho_{\rm T0}/\rho_{\rm t0}=0.35$&$\rho_{\rm T0}/\rho_{\rm t0}=0.64$\\
&$\rho_{\rm T0}/\rho_{\rm EC0}=11.60$&$\rho_{\rm T0}/\rho_{\rm EC0}=10.60$\\
&$\sigma_{\rm T}/\sigma_{\rm t}=2.60$&$\sigma_{\rm T}/\sigma_{\rm t}=2.69$\\
&$\sigma_{\rm T}/\sigma_{\rm EC}=0.44$&$\sigma_{\rm T}/\sigma_{\rm EC}=0.43$\\
&$\Sigma_{\rm T}/\Sigma_{\rm t}=1.25$&$\Sigma_{\rm T}/\Sigma_{\rm t}=2.25$\\
&$\Sigma_{\rm T}/\Sigma_{\rm EC}=3.13$&$\Sigma_{\rm T}/\Sigma_{\rm EC}=2.94$\\
&$z_{\rm t}=110$\,pc&$z_{\rm t}=100$\,pc\\
&$z_{\rm T}=520$\,pc&$z_{\rm T}=500$\,pc\\
&$z_{\rm EC}=2540$\,pc&$z_{\rm EC}=2470$\,pc\\
\hline
\multirow{11}{*}{$0.5\,r_{25}<R<0.8\,r_{25}$}&\multicolumn{2}{c}{$\mu_{\rm l}=25.67\,{\rm mag\,arcsec^{-2}}$}\\
&\multicolumn{2}{c}{$\Delta m=5.5\,{\rm mag\,arcsec^{-2}}$}\\
&$\rho_{\rm T0}/\rho_{\rm t0}=0.39$&$\rho_{\rm T0}/\rho_{\rm t0}=0.70$\\
&$\rho_{\rm T0}/\rho_{\rm EC0}=5.57$&$\rho_{\rm T0}/\rho_{\rm EC0}=5.00$\\
&$\sigma_{\rm T}/\sigma_{\rm t}=2.35$&$\sigma_{\rm T}/\sigma_{\rm t}=2.50$\\
&$\sigma_{\rm T}/\sigma_{\rm EC}=0.44$&$\sigma_{\rm T}/\sigma_{\rm EC}=0.44$\\
&$\Sigma_{\rm T}/\Sigma_{\rm t}=1.19$&$\Sigma_{\rm T}/\Sigma_{\rm t}=2.19$\\
&$\Sigma_{\rm T}/\Sigma_{\rm EC}=1.58$&$\Sigma_{\rm T}/\Sigma_{\rm EC}=1.48$\\
&$z_{\rm t}=150$\,pc&$z_{\rm t}=150$\,pc\\
&$z_{\rm T}=620$\,pc&$z_{\rm T}=590$\,pc\\
&$z_{\rm EC}=2860$\,pc&$z_{\rm EC}=2700$\,pc\\
\hline
\end{tabular}
\end{center}
Note.~--$\Upsilon_{\rm T}/\Upsilon_{\rm t}$: ratio of the thick and thin disk mass-to-light ratios. $\mu_{\rm l}$: limiting magnitude of the fit. $\Delta m$: dynamical range over which the fit has been produced. $\rho_{\rm T0}/\rho_{\rm t0}$, $\rho_{\rm T0}/\rho_{\rm EC0}$: mid-plane thick to thin and thick to EC mass density ratios. $\sigma_{\rm T}/\sigma_{\rm t}$, $\sigma_{\rm T}/\sigma_{\rm EC}$: thick to thin and thick to EC vertical velocity dispersion ratios. $\Sigma_{\rm T}/\Sigma_{\rm t}$, $\Sigma_{\rm T}/\Sigma_{\rm EC}$: thick to thin and thick to EC stellar column mass density ratios. $z_{\rm t}$, $z_{\rm T}$, $z_{\rm EC}$: thin, thick and EC scaleheights.
\end{table}

\section{Results}

The addition of a third stellar component allows us to fit the surface brightness profile down to $\mu({\rm AB})=26.0\,{\rm mag\,arcsec^{-2}}$. The fit is significantly better than that published in CO11b (Fig.~\ref{fits} and Table~\ref{tablefit}) in the sense that it goes far deeper. The fact that for $0.2\,r_{25}<|R|<0.5\,r_{25}$ $\Delta m$ is $3.0\,{\rm mag\,arcsec^{-2}}$ larger than what was achieved in CO11b and that the outermost fitted $z$ goes from $25\arcsec-45\arcsec$ in CO11b to $70\arcsec-100\arcsec$ in the present work argues for the correctness of the fit, and thus for the physical reality of the third component. The numbers exclude a possible alternative, namely that the better fit is achieved merely by the addition of more free parameters.

The fitted properties of the thin and the thick disk are fairly similar to those reported in CO11b (Table~\ref{tablefit}), and the ratio of the column mass densities $\Sigma_{\rm T}/\Sigma_{\rm t}$ is equal, within 10\%, to that obtained for the three bins for which we succeeded to obtain a fit in CO11b. The mid-plane stellar densities are not much affected by the inclusion of the EC, and its main effect is to reduce the scaleheights of the thin and thick disks. The small effect introduced by the EC in $\Sigma_{\rm T}/\Sigma_{\rm t}$ is due to the fact that the mid-plane density of the EC is $\sim10$ ($\sim10$) times smaller than that of the thick disk and $\sim20$ ($\sim10$) times smaller than that of the thin disk for $\Upsilon_{\rm T}/\Upsilon_{\rm t}=1.2$ ($\Upsilon_{\rm T}/\Upsilon_{\rm t}=2.4$). The fact that possible undetected ECs do not have much influence on the results in CO11b shows the robustness of our fitting approach in general.

Each of the three stellar components has a significant range in vertical height for which it dominates the luminosity profile, making it easy to determine its properties and degeneracies unlikely. The thin disk luminosity dominates for $z<5\arcsec$, the thick disk dominates for $5\arcsec<z<20\arcsec$ and the EC dominates for $z>20\arcsec$. The only problem in the fit is that the thin disk scaleheight is poorly constrained due to poor sampling ($z_{\rm t}$ is around 1.5\arcsec, which is smaller than the FWHM).

The EC, if considered to be a disk, has an exceptionally large scaleheight, $z_{\rm EC}\sim3$\,kpc, comparable only to that of thick disks in NGC~0678, NGC~4437 and NGC~4565 in CO11b. The scaleheight is significantly larger for the $-0.8\,r_{25}<R<-0.5\,r_{25}$ bin probably because its luminosity profile is affected by the brightest loop of the tidal stream (MD09). The $\rho_{\rm T0}/\rho_{EC0}$ ratio for $0.2\,r_{25}<|R|<0.5\,r_{25}$ is significantly higher (a factor of two) than it is for $0.5\,r_{25}<|R|<0.8\,r_{25}$ implying that the scalelength is significantly larger than that of the thin and the thick disk.

Using the weightings in Eq.~5 from CO11b we find that the ratios of masses of the stellar components are $M_{\rm T}/M_{\rm t}=1.22$ ($M_{\rm T}/M_{\rm t}=2.19$) and $M_{\rm T}/M_{\rm EC}=2.74$ ($M_{\rm T}/M_{\rm EC}=2.69$) for $\Upsilon_{\rm T}/\Upsilon_{\rm t}=1.2$ ($\Upsilon_{\rm T}/\Upsilon_{\rm t}=2.4$). When taking the mass of the thick disk and the EC together $\left(M_{\rm T}+M_{\rm EC}\right)/M_{\rm t}=1.67$ ($\left(M_{\rm T}+M_{\rm EC}\right)/M_{\rm t}=3.00$). With a circular velocity speed of $v_{\rm c}=181.7\,{\rm km\,s^{-1}}$ (HyperLEDA), NGC~4013 would not fit in the $M_{\rm T}/M_{\rm t}-v_{\rm c}$ relationship discovered by Yoachim \& Dalcanton (2006) and shown in Fig.~12 of CO11b both when considering  $M_{\rm T}/M_{\rm t}$ and $\left(M_{\rm T}+M_{\rm EC}\right)/M_{\rm t}$. Furthermore, if we consider the thick disk as part of the thin disk and the EC to be the only thick disk in NGC~4013, we get $M_{\rm EC}/\left(M_{\rm t}+M_{\rm T}\right)=0.20$ ($M_{\rm EC}/\left(M_{\rm t}+M_{\rm T}\right)=0.26$), which is too low to fit into the $M_{\rm T}/M_{\rm t}-v_{\rm c}$ relationship.

The luminosity profiles could not be satisfactorily fitted by a sum of two ${\rm sech}^2(z/z_0)$ functions, but were acceptably fitted by a sum of two exponential functions. The $\sqrt{(\chi^2)}$ was, however, worse than in the three disk fit and misses a substantial amount of light corresponding to the thinner component of the three-disk fit ($0.2-0.3\,{\rm mag\,arcsec^{-2}}$ for the inner bins). This fit, although simpler than that made with three coupled stellar components, should be regarded as unphysical.

\section{Discussion}

Just as causal links have been sought between the ``prodigious'' warp and the merging event causing the ``giant'' stream, one may also be tempted to find links between a minor merger and the EC of NGC~4013.

\begin{figure}
\begin{center}
\begin{tabular}{c}
\includegraphics[width=0.45\textwidth]{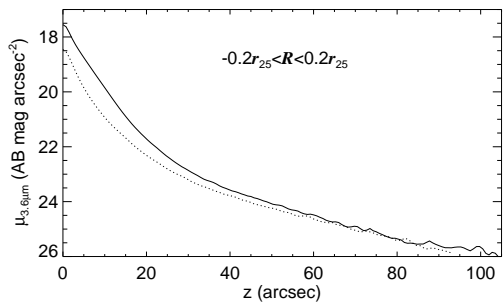}\\
\end{tabular}
\caption{\label{centralbin} Luminosity profile for the $-0.2\,r_{25}<R<0.2\,r_{25}$ bin (solid line) compared to that obtained from the $0.2\,r_{25}<|R|<0.5\,r_{25}$ bins (dotted line).}
\end{center}
\end{figure}

If the EC is linked to the minor merger, then it can be made of tidal debris. According to the images in MD09, the tidal loops should not affect the luminosity profiles at $R=0$. Although producing a fit for the minor axis using our procedure is impossible due to the presence of the bulge, we produced the luminosity profile (Fig.~\ref{centralbin}). It is clear that in the bin with $-0.2\,r_{25}<R<0.2\,r_{25}$ the EC is present and probably slightly brighter than in the $0.2\,r_{25}<|R|<0.5\,r_{25}$ as predicted in the case of a disk or an ellipsoid. The color analysis made by MD09 shows that the tidal stream is much redder than the EC, although the large error bars make a common origin still possible. Other arguments against a tidal origin for the EC are its smoothness, its symmetry and the uniformity of its scaleheight with varying radial distance $R$.

Another possibility is that the EC has been formed by the dynamical heating of the disks by the crossing of the dwarf galaxy causing the tidal stream. However, as the disk self-gravity which counteracts the disk heating is higher at low galactocentric radii, disks formed in this way appear flared and, as a consequence, the EC would form with a relatively small scaleheight at low $R$. Simulations (Quinn et al.~1993; Walker et al.~1996; Kazantzidis et al.~2008; Bournaud et al.~2009) show how this mechanism could reasonably produce the observed scaleheights at high galactocentric radius, but would fail to produce such an EC for the inner kiloparsecs. Additionally, it seems difficult to create a component containing 20\% (26\%) of the mass of the galaxy if we assume $\Upsilon_{\rm T}/\Upsilon_{\rm t}=1.2$ ($\Upsilon_{\rm T}/\Upsilon_{\rm t}=2.4$) in a recent event without disturbing the disks. The regularity and the symmetry of the galaxy argues for an old origin for the EC, thus confirming that the assumption of equilibrium is reasonable. Finally, the stream in NGC~4013 is Monoceros-like (MD09), thus probably a very minor merger ($\sim$1:100; Pe\~narrubia et al.~2005) not likely to cause large disturbances. The possibility of the EC being created during an older merger event cannot be discarded; Purcell et al.~(2010) have shown that a 1:10 merger 5\,Gyr ago in a Milky Way-like galaxy yields results which qualitatively match the appearance of NGC~4013 except for the fact that their resulting EC is more boxy than the one we detect and we do not find the same flaring. They show that low-latitude accretion events could heat up to 1\% of the disk mass to $z>7$\,kpc, which is in rough agreement with what we found ($1.8\%$ of disk mass at $z>7$\,kpc).

We also discard the EC to be the effect of a stellar warp seen in projection. Bottema (1995) found that the stellar disk is truncated at the radius at which the warp starts and the warp model made by Bottema (1996) shows that its line of nodes is in the direction of the line of sight, making a stellar warp look as a thick disk when seen in projection impossible. Even if the \hi\ warp and the stellar warp were decoupled and with different lines of nodes the stellar warp would be unlikely to mimic an EC unless the stellar warp started at a radius smaller than the \hi\ warp.

Thus, we deduce from our findings and from literature data that the EC is a real feature and not some tidal feature or warp seen in projection. We also deduce that both the thick disk and the EC are old. A possible scenario for the formation of a three component system would be having a disk formed thick with stars formed before and during the build-up of the galaxy from small fragments (Robertson et al.~2006; Brook et al.~2007; Richards et al.~2010) and/or a disk heated by the internal thickening caused by kinematical heating due to giant clumps (Elmegreen \& Elmegreen 2006). Then a merger event would further thicken the disk, which would become what we know as the EC. After the merger, the canonical thick disk would form with the same mechanisms as the EC formed prior to the merger event. Finally, the remaining gas, plus that coming from cold flows would settle in the mid-plane and form the thin disk. This disk formation mechanism is not very frequent, since only two galaxies over 30 exhibit a bright EC in CO11b. The difference in the formation mechanisms may explain why these two galaxies fall outside the $M_{\rm T}/M_{\rm t}-v_{\rm c}$ relationship discovered by Yoachim \& Dalcanton (2006).

\section{Conclusions}

NGC~4013 is not adequately described by the canonical thin+thick disk description (CO11b). In this Letter, the luminosity profiles of NGC~4013 are fitted satisfactorily using the solutions of three stellar flattened components plus one gaseous disk in equilibrium. The newly described extended component (EC) has a relatively low surface brightness, but due to its vertical extent, contains a significant fraction of the disk mass (between 20\% and 26\% depending on the assumed $\Upsilon_{\rm t}/\Upsilon_{\rm T}$). The EC has a longer scalelength than the galaxy disks, is smooth and its properties do not depend strongly on varying galactocentric radii.

The nature of the EC is unknown and could be a second thick disk or a squashed elliptical halo. The smoothness of the EC makes it unlikely to be related to the ongoing minor merger of NGC~4013. We also discard the EC to be made of off-plane stars of a warped disk, since the warp has been modeled to have its line of nodes in the direction of the line of sight (Bottema 1996). We favor a scenario in which the EC was formed in a two-stage process, in which an initially thick disk was dynamically heated by a merger soon enough in the galaxy history to have a new thick disk formed within it.

\section*{Acknowledgments}

The authors wish to thank the entire S$^4$G team for their efforts in this project. This work is based on observations and archival data made with the Spitzer Space Telescope, which is operated by the Jet Propulsion Laboratory, California Institute of Technology under a contract with NASA. We are grateful to the dedicated staff at the Spitzer Science Center for their help and support in planning and execution of this Exploration Science program. We gratefully acknowledge support from NASA JPL/Spitzer grant RSA 1374189 provided for the S$^4$G project. EA and AB thank the CNES for support. KS, J-CMM, TKim and TMizusawa acknowledge support from the National Radio Astronomy Observatory, which is a facility of the National Science Foundation operated under cooperative agreement by Associated Universities, Inc. Funding for the SDSS has been provided by the Alfred P.~Sloan Foundation, the Participating Institutions, the National Science Foundation, the U.S. Department of Energy, NASA, the Japanese Monbukagakusho, the Max Planck Society, and the Higher Education Funding Council for England. This research has made use of the NASA/IPAC Extragalactic Database (NED) which is operated by JPL, CALTECH, under contract with NASA.

\end{document}